\documentclass[lettersize,journal]{IEEEtran}
\usepackage{amsmath,amsfonts}
\usepackage{algorithmic}
\usepackage{algorithm}
\usepackage{array}
\usepackage[caption=false,font=normalsize,labelfont=sf,textfont=sf]{subfig}
\usepackage{textcomp}
\usepackage{stfloats}
\usepackage{url}
\usepackage{verbatim}
\usepackage{graphicx}
\usepackage{cite}
\usepackage{float}
\usepackage{color,soul}
\newcommand\three{3$\times$3-mm}
\newcommand\six{6$\times$6-mm}
\newcommand\threea{3$\times$3-mm OCT angiogram}
\newcommand\sixa{6$\times$6-mm OCT angiogram}
\newcommand\threeas{3$\times$3-mm OCT angiograms}
\newcommand\sixas{6$\times$6-mm OCT angiograms}

\newcommand\loss{\mathcal{L}}
\newcommand\down{$\downarrow$}
\newcommand\up{$\uparrow$}
\usepackage{amsmath}
\usepackage{amssymb,amsfonts}
\usepackage{pifont}
\usepackage{tikz}
\newcommand{\tikzxmark}{%
\tikz[scale=0.23] {
    \draw[line width=0.7,line cap=round] (0,0) to [bend left=6] (1,1);
    \draw[line width=0.7,line cap=round] (0.2,0.95) to [bend right=3] (0.8,0.05);
}}

\usepackage[normalem]{ulem}
\useunder{\uline}{\ul}{}
\usepackage{caption}
\usepackage{xcolor,colortbl}
\usepackage{hyperref, bookmark}						
\usepackage{bbding, makecell} 
\usepackage{array}
\usepackage{booktabs} 
\usepackage{multirow}

\begin{document}

\title{Reference-based OCT Angiogram Super-resolution with Learnable Texture Generation}

\author{Yuyan Ruan, \IEEEmembership{Student Member, IEEE}, Dawei Yang, Ziqi Tang, An Ran Ran, Carol Y. Cheung and

Hao Chen, \IEEEmembership{Senior Member, IEEE}


\thanks{This work was supported by funding from Center for Aging Science, Hong Kong University of Science and Technology, and Shenzhen Science and Technology Innovation Committee (Project No. SGDX20210823103201011), and Direct Grants from The Chinese University of Hong Kong (Project Code: 4054419 \& 4054487). (Corresponding author: Hao Chen.)}
\thanks{Yuyan Ruan is with the Department of Chemical and Biological Engineering, The Hong Kong University of Science and
Technology, Clearwater Bay, Hong Kong (e-mail: yruanaf@connect.ust.hk).}
\thanks{Dawei Yang, Ziqi Tang, An Ran Ran, Carol Y. Cheung are with Department of Ophthalmology and Visual Sciences, The Chinese University of Hong Kong, Hong Kong.}
\thanks{Hao Chen is with the Department of Computer Science and Engineering and the Department of Chemical and Biological Engineering, The Hong Kong University of Science and
Technology, Clearwater Bay, Hong Kong (e-mail: jhc@cse.ust.hk).}
}
\markboth{IEEE TRANSACTIONS ON ***}%
{RUAN \MakeLowercase{\textit{et al.}}: Reference-based OCT Angiogram Super-resolution with Learnable Texture Generation}



\maketitle

\begin{abstract}
Optical coherence tomography angiography (OCTA) is a new imaging modality to visualize retinal microvasculature and has been readily adopted in clinics. High-resolution OCT angiograms are important to qualitatively and quantitatively identify potential biomarkers for different retinal diseases accurately. However, one significant problem of OCTA is the inevitable decrease in resolution when increasing the field-of-view given a fixed acquisition time. To address this issue, we propose a novel reference-based super-resolution (RefSR) framework to preserve the resolution of the OCT angiograms while increasing the scanning area. Specifically, textures from the normal RefSR pipeline are used to train a learnable texture generator (LTG), which is designed to generate textures according to the input. The key difference between the proposed method and traditional RefSR models is that the textures used during inference are generated by the LTG instead of being searched from a single reference image. Since the LTG is optimized throughout the whole training process, the available texture space is significantly enlarged and no longer limited to a single reference image, but extends to all textures contained in the training samples. Moreover, our proposed LTGNet does not require a reference image at the inference phase, thereby becoming invulnerable to the selection of the reference image. Both experimental and visual results show that LTGNet has superior performance and robustness over state-of-the-art methods, indicating good reliability and promise in real-life deployment. The source code will be made available upon acceptance.
\end{abstract}

\begin{IEEEkeywords}
Deep learning, Optical coherence tomography angiography, Reference-based super-resolution, Retina, Texture generation
\end{IEEEkeywords}

\section{Introduction}
\label{sec:introduction}
\begin{figure*}[t]
\centering
\includegraphics[width=0.9\textwidth]{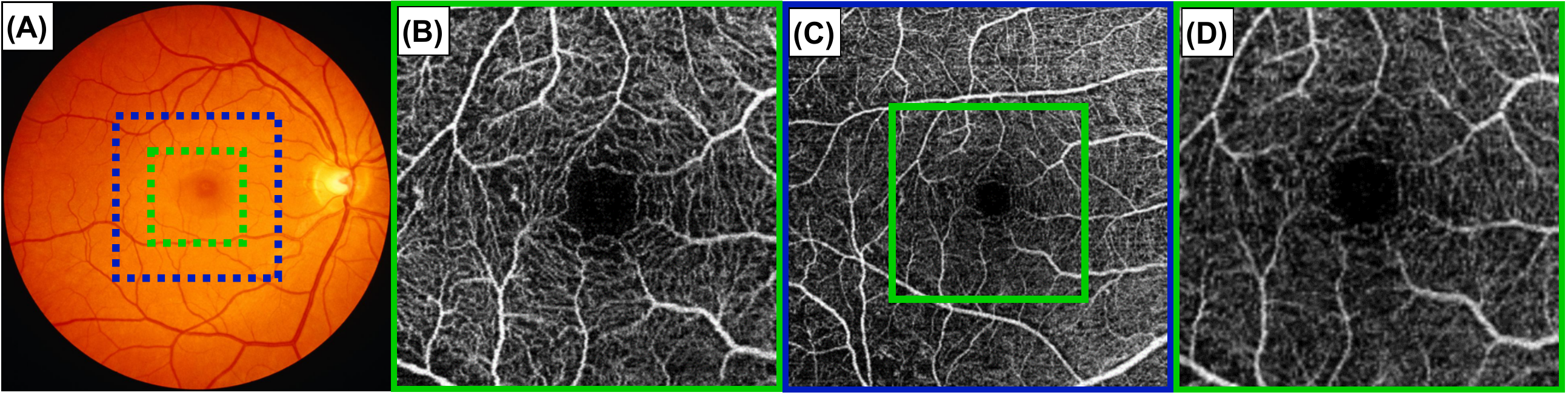}
\caption{Illustration of spatial relationship between \three\ and \sixas.}
\label{legend}
\end{figure*}
Optical coherence tomography angiography (OCTA)  is a novel modality that can provide high-resolution and depth-resolved angiographic flow images to visualize retinal microvasculature. With  the advantage of non-invasiveness and readiness, it has been widely used to study retinal abnormalities and diseases such as diabetic macular ischemia and choroidal neovascularization \cite{octa2020review, dmi2022gabriel1, dmi2022gabriel2_cohort2022}. 

Despite the great advantages, there are still some drawbacks of this technique, and one of them is the resolution drop when increasing the field-of-view (FOV). 
The mechanism behind is that the number of B-scans is fixed regardless of the scanning area due to the nature of OCTA imaging technique. Therefore, the B-scan density will drop as the FOV increases. 
The relationship between $3\times 3-$ and \sixas\ is shown in Fig. \ref{legend}. (A) shows a photo of the retina, where the green and blue dotted boxes annotated the region covered by $3\times 3$ and \sixas, respectively. (B) and (C) show a $3\times 3$- and \sixas\. The region covered by \threeas\ is annotated by the green box in (C). (D) is obtained by cropping out the patch annotated by the green box in (C), and then upsampled by a factor of 2. It can be observed that the clarity of (B) and (D) is very different despite the fact that they cover the same physical region due to different scanning densities. 
The scanning density under a small FOV like \threeas, is sufficient to clearly present microvasculature and delineate the foveal avascular zone (FAZ) \cite{octa3}.  
Contrastively, the detailed capillaries may be missing or corrupted in scans with larger FOVs, like \sixas\, since the resolution is significantly lower. However, they are still important in clinical practices because of their broader coverage, which is helpful in detecting more pathological features such as microaneurysms and non-perfusion \cite{octa2020wide, rose2020, largefov}.
Consequently, OCT angiograms with both large FOVs and high resolution are strongly desired \cite{dawei2023review}. Unfortunately, such images could be difficult to obtain directly since the acquisition time might be increased multiple times under the same scanning density. A longer acquisition time places higher demand on patients' tolerance, and the resultant images are prone to various artifacts caused by unintentional motions such as blinking and eye movements \cite{largefov_artifacts}. As an alternative to increasing the acquisition time, image enhancement techniques like image super-resolution can be applied to improve the resolution of OCT angiograms.

Image super-resolution (SR) is an image processing technique to recover the high-resolution (HR) images given their low-resolution (LR) version, which can be applied to improve the resolution of angiograms with large FOVs. However, only a few works have targeted using them to enhance OCT angiograms despite the rapid development of deep SR networks. Existing methods for OCTA SR task solely rely on the information contained in the input LR images and have not utilized information from the HR domain. For another thing, there emerges a new category of SR approach named as reference-based super-resolution (RefSR) which makes use of extra information from the target domain to assist the SR process. These approaches have demonstrated extraordinary power in SR task benefiting from the information of the target domain. 

In such a context, we propose a novel RefSR model to address the resolution degradation issue in OCT angiograms due to inadequate sampling. Based on the idea of RefSR, a learnable texture generator (LTG) is introduced to enlarge the texture space as well as mitigate the influence of the reference image selection.
Concretely, the overall proposed LTGNet replaces textures searched from the reference (Ref) images with those generated by the LTG. By training the LTG with plenty of LR-Ref pairs, the generated textures are thereby not limited to the single input Ref image. Moreover, the model robustness is enhanced since the performance is no longer affected by the selection of Ref images. The proposed method showed superior and robust performance over the state-of-the-art approaches during the experiments. In addition, since the common scanning areas of OCTA are 3$\times$3- or \six, ground truth images of the \sixas\ are usually only available for their central part. Therefore, it is important to ensure stable performance on the entire \six\ area when the model is trained only on the center overlapped areas of $3\times3$ and \sixas. To the best of our knowledge, we are the first to validate the model performance on the whole \six\ area. 

In summary, we proposed a novel RefSR model to obtain OCT angiograms with both large FOVs and high resolution, which also demonstrates robust performance. We further validated our model on the whole 6$\times$6-mm area to ensure clinical reliability. 
\section{Related Work}
\subsection{Image Super-resolution}
Image super-resolution (SR) refers to reproducing the high-resolution (HR) image when given its low-resolution (LR) counterpart. 
Current SR methods can be divided into two main streams: single image super-resolution (SISR) and reference-based image super-resolution (RefSR). In SISR, the networks try to reconstruct the HR image from the information contained in the input LR image and learn the mapping between it and its corresponding HR version from extensive training examples \cite{sisr2019deep}. The first deep learning based SR model, SRCNN, was proposed by \cite{srcnn2014} in 2014. Later on, deeper and more powerful models were proposed to improve the performance. 
VDSR \cite{vdsr2016} introduced residual learning to effectively simplify the SR process, which enabled stably training much deeper network with a larger learning rate. DenseNet \cite{dense2017} further exploited dense skip connections to combine both low- and high-level features and achieved significantly better performance. 
However, models trained by minimizing the pixel-wise mean squared error (MSE) between the HR and LR images usually result in blurry details and fail to output plausible results. To address this issue, SRGAN \cite{srgan} introduced generative adversarial networks (GANs) as well as perceptual loss into the SR domain to produce more realistic output. 

Unfortunately, the capability of SISR is limited since high-frequency details are usually hard to recover directly from LR images. Therefore, RefSR has been proposed to resolve this issue by exploiting useful textures in the given extra high-resolution reference (Ref) image to facilitate the SR process. With such information from the HR domain, the network is able to handle more challenging tasks, e.g., $\times 8$ upscaling, and produce finer and more plausible details. 
Landmark \cite{landmark2013refsr}, one of the earliest RefSR models, retrieved similar images from the Internet and applied global registration to align the upsampled LR image with the obtained Ref image. A similar paradigm was adopted in CrossNet \cite{crossnet2018refsr}, where optical flow was utilized to do the alignment. However, such methods cannot handle Ref images with great displacement from the LR images. 

Recent models have started to focus on relaxing the constraints that the Ref and LR images should share similar contents. 
SRNTT \cite{srnttzhang2019} proposed a neural texture transfer model that matches and fuses features of the Ref and LR images at patch level to tackle the discrepancy between them. TTSR \cite{ttsr} further improved the performance by using a trainable encoder and was the first to introduce the transformer architecture in the RefSR domain. 
Later on, MASA \cite{masa} proposed a coarse-to-fine searching mechanism and reduced the computational cost of texture searching significantly. Yet such matching and transfering frameworks are still suboptimal since the available texture space is limited to one single Ref image, and the performance is vulnerable to the selection of Ref images. \cite{refpool2020ecva} proposed to construct a reference pool using Farthest Point Sampling and to search for the best matched texture features during testing. This method enlarges the available texture space to some extent but introduces extra storage and a longer searching time. In contrast, our method replaces the static reference pool with a learnable texture generator, which can avoid the storage of texture features and the extra searching costs.

\subsection{OCT Angiogram Super-resolution}
Image super-resolution for OCT angiograms with large FOVs is gaining attention. HARNet \cite{harnet2020reconstruction} was the first method to be proposed for the super-resolution of the superficial capillary plexus (SCP) of OCT angiograms, followed by DCARNet \cite{dcarnet2021} designated for intermediate capillary plexus (ICP) and deep capillary plexus (DCP). These two models are essentially composed of stacks of convolutional layers and residual blocks and optimized by a loss function with regard to both MSE and structural similarity (SSIM). Although they have produced satisfying results, the performance is still limited by the network capability. \cite{octasargan2022} devised a generative adversarial network (GAN) and obtained perceptually better results. \cite{octasparse2022} further incorporated the idea of domain adaptation to tackle the discrepancy between the LR and HR domains. Furthermore, they also put forward a sparse edge-aware loss to refine the vessel structure. For another thing, \cite{weiwen2022freq} proposed an unsupervised model and exploited the frequency information for better reconstruction results. 

Despite the recent development, existing methods merely rely on the information contained in the input LR angiograms and have not utilized information from the HR domain, which still limits the model performance. Therefore, incorporating information from the target domain into the OCTA SR problem remains underexplored and is worth investigating. In this context, we put forward a novel RefSR method for the super-resolution of OCT angiograms, which can incorporate information from both LR inputs and some high-frequency texture information from the HR domain. Experimental results indicate that our model has superior performance over state-of-the-art methods for not only the foveal-central \three\ area but also the whole \six\ area when only trained with the central parts, showing great robustness and generalisability.

\section{Methodology}
\begin{figure*}[t]
\centering
\includegraphics[width=0.9\textwidth]{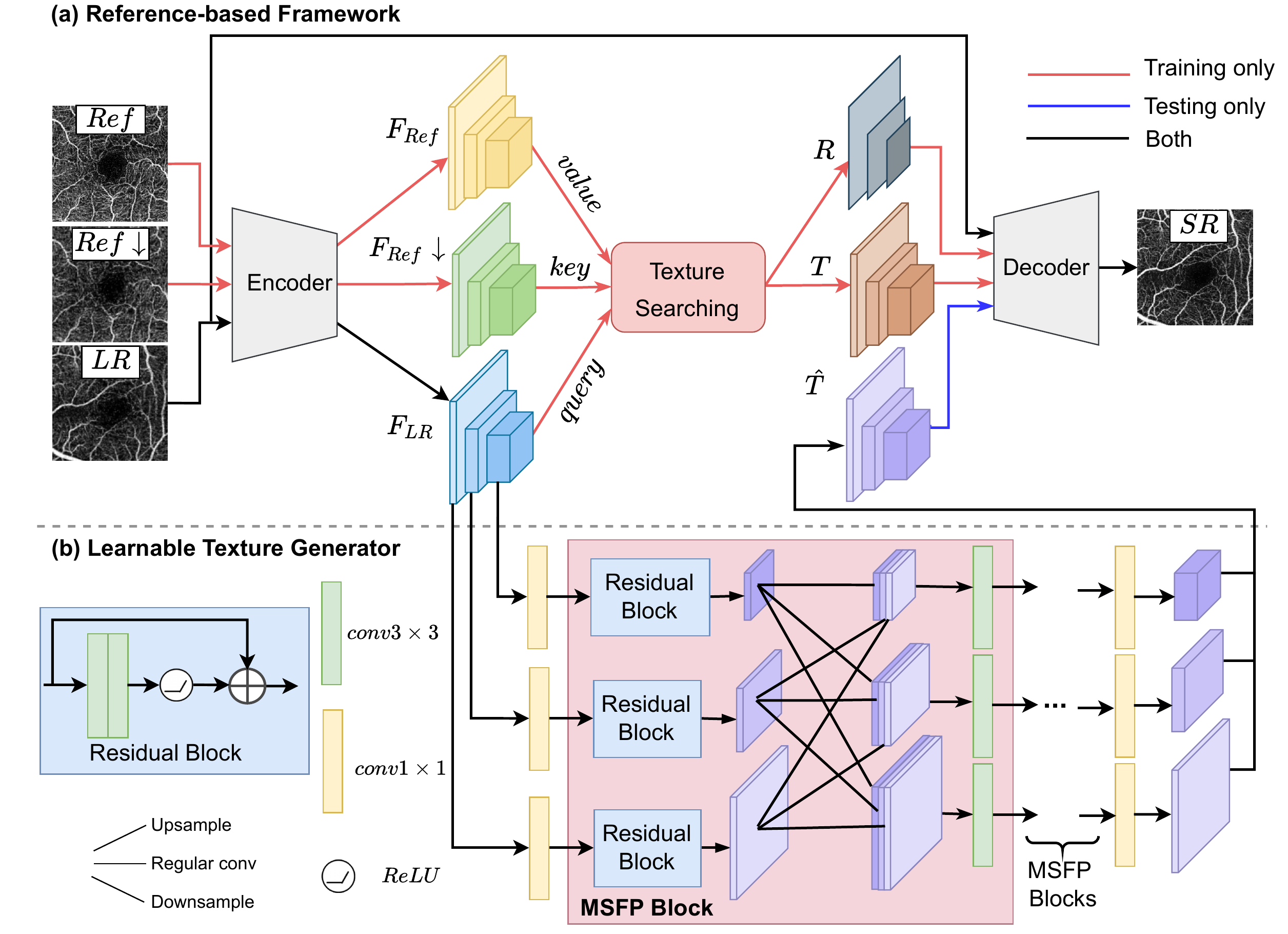}
\caption{Overview of the LTGNet. (a) shows the overall reference-based framework. (b) depicts the detailed structure of the LTG. Paths involved at the various stages are marked by different colors. }
\label{model}
\end{figure*}

The proposed LTGNet is shown in Fig. \ref{model}. It consists of a normal RefSR model and an LTG. The RefSR model includes essential building blocks like an encoder to extract features from images, texture searching module and a decoder to output the final SR image. During training, textures extracted by the RefSR model are used as ground truth to train the LTG, while at the inference stage, textures used in the decoder are generated by the LTG instead of being searched from the Ref image. Each component will be elaborated in detail in the following subsections. 

\subsection{Reference-based Framework}
\begin{figure}[H]
\centering
\includegraphics[width=0.48\textwidth]{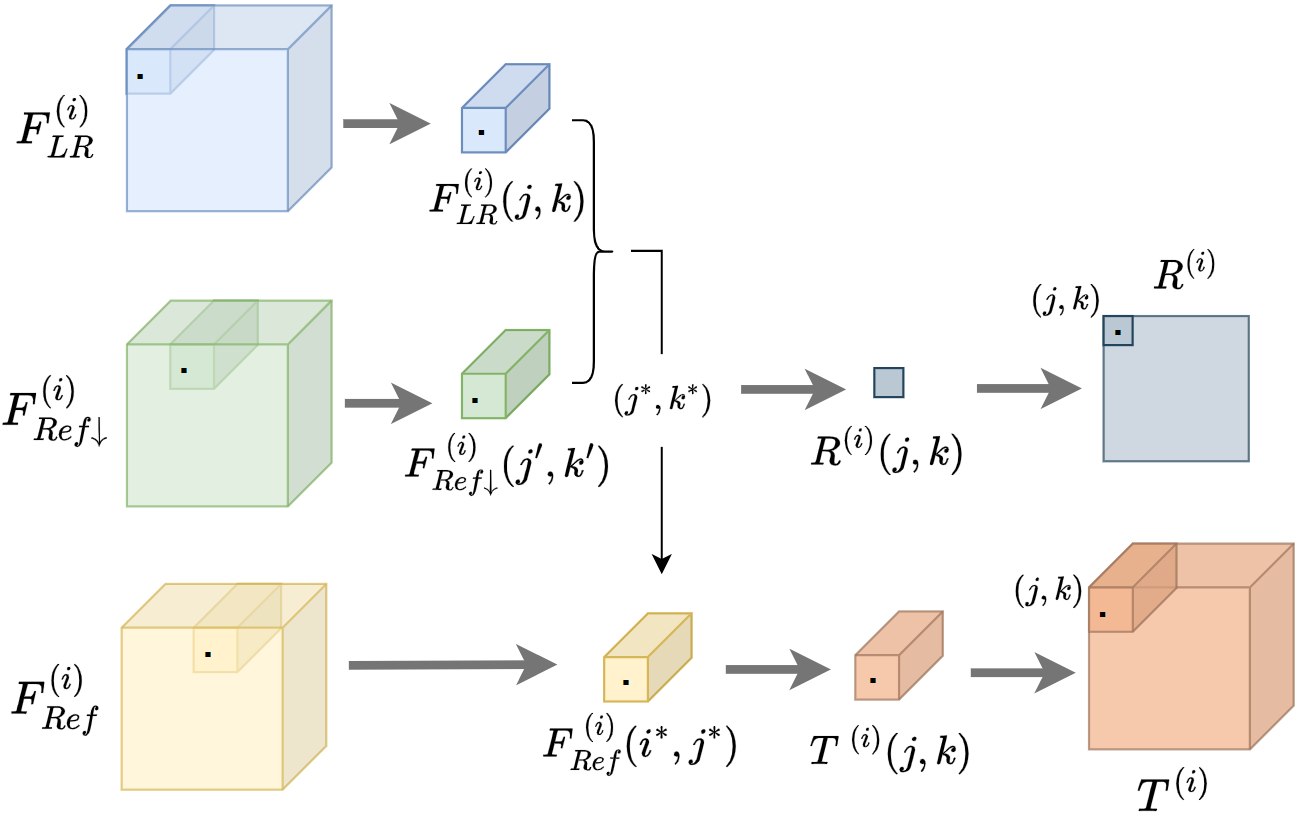}
\caption{Illustration of the texture searching process.}
\label{search}
\end{figure}

Since it is hard to directly recover missing high-frequency details from LR images, a RefSR model is adopted to assist the SR procedure by extracting proper textures from the high-resolution Ref image. After identifying similar patterns in the LR and Ref pair, the model can transfer and merge the high-resolution textures from the Ref image with the LR input. 
The texture searching procedure is illustrated in Fig. \ref{search}. This texture searching process is conducted at multiple scales to extract both shallow as well as deep texture features. Here, we illustrate the process with respect to scale $i$.
Since the distributions of the LR images and the high-resolution Ref images are quite different, directly calculating the similarity between them is not reasonable. Therefore, a degraded version of Ref image (Ref$\downarrow$) will be obtained first. In our case, Ref$\downarrow$ is the corresponding \sixa\ of the Ref image. 
After that, $F_{LR}^{(i)}$ is taken as the query, $F_{Ref\downarrow}^{(i)}$ as the key and $F_{Ref}^{(i)}$ as the value in the attention mechanism \cite{att2017,att2021}. Here, $F_{*}^{(i)}$ represents features of image * extracted by the encoder at scale $i$. 
To be specific, $F_{LR}^{(i)}$, $F_{Ref\downarrow}^{(i)}$ and $F_{Ref}^{(i)}$ are first unfolded into $p\times p \times C$ patches, where $p$ is the patch size and $C$ is the number of channels of the feature maps.
The inner product between two patches is used to measure the relevance.
For every LR feature patch $F_{LR}^{(i)}(j,k)$, the most relevant patch in $F_{Ref\downarrow}^{(i)}$ is searched and denoted as $F_{Ref\downarrow}^{(i)}(j^*,k^*)$, where $F_{*}^{(i)}(j,k)$ represents the patch of $F_*^{(i)}$ centered at $(j,k)$.
The returned textures $T^{(i)}$ and the relevance map $R^{(i)}$ are then formed by the following operations:

\begin{equation}
    R^{(i)}(j,k) = \left \langle F_{LR}^{(i)}(j,k),F_{Ref\downarrow}^{(i)}(j^*,k^*)\right \rangle
\end{equation}
\begin{equation}
    T^{(i)}(j,k) = F_{Ref}^{(i)}(j^*,k^*),
\end{equation}
where $\langle\ \rangle$ denotes the inner product.
Finally, the textures $T$ and the relevance map $R$, together with the LR features $F_{LR}$, will be further rendered into the final SR image by the decoder. 
\subsection{Learnable Texture Generator}
The LTG is proposed to solve the limited texture space problem in conventional RefSR methods. The LTG is trainable and optimized by plenty of LR-Ref pairs at the training stage. As a result, the textures generated at the inference phase are not limited to the given single Ref image but from all the textures that the model has learned at the training stage. Moreover, the model performance will not be affected by the selection of Ref images in this way since no Ref image is required at inference time.
Considering textures are not only related to superficial and local features but also semantic meanings, the LTG is designed to integrate LR features at both the low- and high-level to generate more appropriate and powerful texture representations.

As shown in Fig. \ref{model} (b), the LTG takes LR features of multiple scales as input and generates some useful textures accordingly.
Firstly, the LR features  are processed by a $1\times 1$ convolution layer before going into a series of multi-scale feature processing (MSFP) blocks. Each MSFP block is composed of one residual block and a multi-scale feature fusion module \cite{msff}, where features of different scales are upsampled or downsampled to the same size and stacked together before being merged by another convolution layer. 
In this way, information of multiple scales is exchanged and integrated to generate more powerful feature representations. Lastly, each feature block is processed by a $1\times 1$ convolutional layer to produce the final generated textures $\hat{T}$ with the desired channel numbers. Since the shape of the features remains unchanged after each MSFP block, such blocks can be easily stacked in series according to need. 
\subsection{Hybird Feature Fusion Decoder}
\begin{figure}[t]

\includegraphics[width=0.45\textwidth]{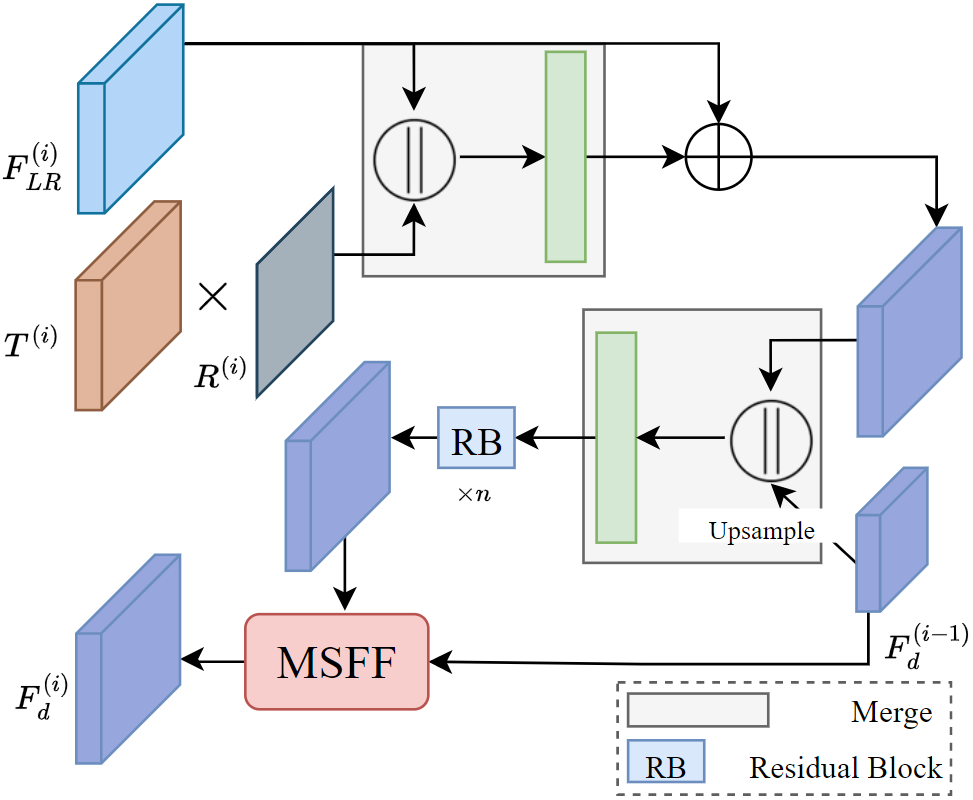} 
\caption{Basic building block of the decoder.}
\label{decoder}
\end{figure}

The decoder fuses LR features and textures progressively from the smallest scale to the largest. Its basic building block is illustrated in Fig. \ref{decoder} and is applicable to every scale in the model. At scale $i$, the texture of the corresponding scale $T^{(i)}$ is multiplied with the relevance map $R^{(i)}$ first so that the texture with higher similarity will receive larger weighting, and vice versa. 
These weighted texture and $F_{LR}^{(i)}$ are then processed by a "merging" operation, which consists of concatenation and a convolutional layer to preserve channel number consistency. A residual connection between $F_{LR}^{(i)}$ and the merged feature is also established to preserve the fidelity of the input feature as well to alleviate the learning difficulty. Then, the decoder feature from the smaller scale $F_d^{(i-1)} $ is upscaled via bicubic-interpolation and "merged" with the feature obtained from the previous steps, which produces the intermediate decoder feature $F_{d'}^{(i)}$. After several residual blocks, $F_{d'}^{(i)}$ and all  $F_{d}^{(j)}$ with $j< i$ are sent into an MSFP block to give the refined decoder feature  $F_{d}^{(i)}$, where information of different scales is exchanged and fused. For the smallest scale ($i=1$), the integration of the feature from the smaller scale and the corresponding MSFP block are omitted since there is no such $F_d^{(i-1)}$. Lastly, the fused and refined decoder features of all scales are reshaped to the target scale and then concatenated before being processed by a convolutional layer to produce the SR image.
\subsection{Overall Loss Function and Optimization}
In order to stablize the training as well as sharpen the output image, we adopt l1-norm loss as reconstruction loss \cite{l1zhu2017,srnttzhang2019}:
\begin{equation}
\loss_{rec}=||HR-SR||_1,
\end{equation}
where the HR is the ground truth high-resolution image.
Perceptual loss $\loss_{per}$ is also included to preserve the perceptual quality of the output SR images \cite{perceptualjohnson2016, perceptual2017enhancenet, perceptuallucas2019}, which can be calculated from the following equation:
\begin{equation}
\begin{aligned}
    \mathcal{L}_{per} =& \frac{1}{C_1\cdot H_1 \cdot W_1}||f(HR)-f(SR)||_{2}^{2} +\\
    & \frac{1}{C_2\cdot H_2 \cdot W_2\cdot k}\sum_{i=1}^{k}||g(SR)^{(i)}-T^{(i)})||_2^2,
\end{aligned}
\end{equation}
where $f$ is the relu5\_1 layer of VGG19 \cite{vgg19simonyan2014very} pre-trained on ImageNet \cite{imgnet} in our case.The first part aims to minimize the difference of the SR and HR images in latent feature space. The second part is to encourage the SR image to have similar texture features to the extracted textures $T$ \cite{ttsr}, and $g(SR)^{(i)}$ represents the features of the SR image extracted by the encoder at scale $i$. The number of scales included in the calculation is denoted as $k$, and the channel number, height and width of the feature/textures maps are represented as $C_*$, $H_*$ and $W_*$, respectively. 
\begin{figure*}[t]
	\centering
	\includegraphics[width=0.9\textwidth]{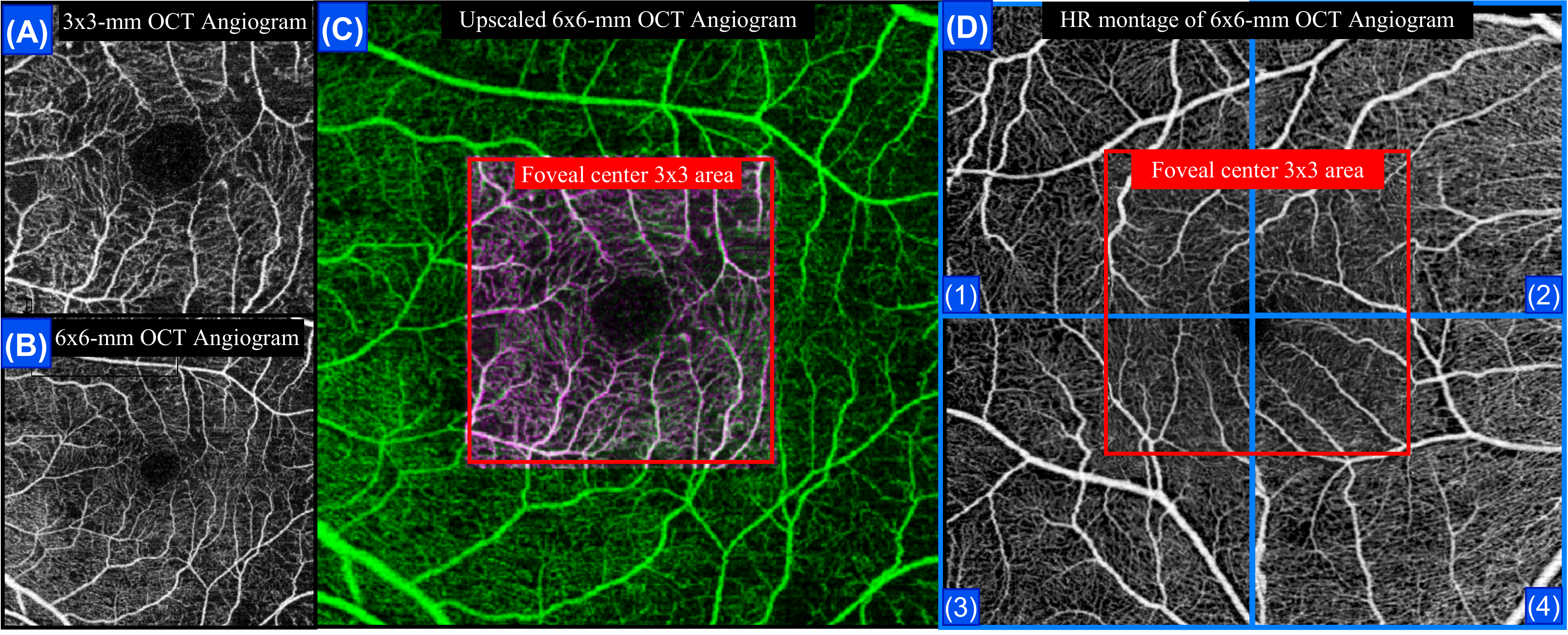}
	\caption{Data illustration and pre-processing steps. (A) and (B) are a pair of $3\times3$ and \sixas. (C) shows the registered image of the upscaled $6\times 6$-mm OCT angiogram and the high-resolution \threea. (D) is the HR montage of the \sixa\ stitched with five \threeas.}
	\label{fig:data}
\end{figure*}
To further improve the visual quality of the generated SR images, we include the adversarial loss $\loss_{adv}$ described in WGAN-GP\cite{wgangp2017}, which can be formulated as
\begin{equation}
    \begin{aligned}
\mathcal{L}_{D}=& \underset{\tilde{x} \sim \mathbb{P}_{g}}{\mathbb{E}}[D(\tilde{x})]-\underset{x \sim \mathbb{P}_{r}}{\mathbb{E}}[D(x)]\\
&+ \lambda \underset{\hat{x} \sim \mathbb{P}_{\hat{x}}}{\mathbb{E}}\left[\left(\left\|\nabla_{\hat{x}} D(\hat{x})\right\|_{2}-1\right)^{2}\right] 
    \end{aligned}
\end{equation}
\begin{equation}
    \begin{aligned}
    \mathcal{L}_{G}=&-\underset{\tilde{x} \sim \mathbb{P}_{g}}{\mathbb{E}}[D(\tilde{x})],
    \end{aligned}
\end{equation}where $G$ and $D$ refer to LTGNet and the extra discriminator, respectively. For $D$, we adopt the same structure as \cite{srnttzhang2019} and \cite{ttsr}.
Lastly, we propose the texture generation loss to train the LTG by enforcing consistency between the generated and searched textures in latent space, which can be formulated as follows. 
\begin{equation}
\mathcal{L}_{tg}=\frac{1}{n} \sum_{i=1}^{n}\frac{1}{C_i\cdot H_i \cdot W_i} \left\|T^{(i)}-\hat{T}^{(i)}\right\|_{2}^2\odot R^{(i)},
\end{equation}
where $\odot$ represents the point-wise multiplication and $n$ is the number of scales that $T$ contains
The overall loss function can be written as:
\begin{equation}
\loss =\loss_{rec}+\lambda_{per}\loss_{per}+\lambda_{tg}\loss_{tg}+\lambda_{adv}\loss_{adv}
\end{equation}
\section{Experiments and Results}
\subsection{Dataset and Preparation}

\subsubsection{Data Collection}
The dataset was retrospectively collected from the Chinese University of Hong Kong Sight-Threatening Diabetic Retinopathy (CUHK-STDR) study, which was an observational clinical study for diabetic retinopathy in subjects with Type 1 or Type 2 Diabetes Mellitus recruited from CUHK Eye Centre, Hong Kong Eye Hospital \cite{cohort2019, dmi2022gabriel2_cohort2022} \footnote{This study involves human participants and was approved by the Hong Kong Kowloon Central Cluster Research Ethics Committee (Kowloon Central/Kowloon East, reference number KC/KE-15-0137/ER-3).}. The OCT angiograms were generated by swept-source optical coherence tomography (OCT) (DRI OCT Triton; Topcon Inc, Tokyo, Japan), where the wavelength, acquisition speed, axial and transversal resolution were set to 1050 nm, 100,000 A-scans/s, 7 and 20 $\mu m$, respectively. Every acquired OCTA volumetric scan was centered at the fovea and comprised 320 B-scans, each of which consisted of 320 A-scans. The 2D OCT angiograms were the maximum projection of the OCTA signal of the superficial capillary plexus, which was referenced to 2.6  $\mu m$ below the internal limiting membrane to 15.6 $\mu m$ below the junction between the inner plexiform and the inner nuclear layers. 
To create the external and comprehensive test set, five \threeas\ (one foveal-centered plus four parafoveal-centered, corresponding to the red square and four blue squares in Fig. \ref{fig:data} D) were purposely acquired to create an HR montage of the \sixa\ \cite{octa3}, providing ground truth for the whole \six\ area. 
\subsubsection{Data Preprocessing}
As shown in Fig. \ref{fig:data} (C), to obtain the LR-HR image pair, the original \sixa\ (B) was upscaled by a factor of 2 using bicubic interpolation before registering it with the paired \threea\ (A). After that, the largest inscribed rectangles of the overlapping areas were cropped out to produce an LR-HR pair. 
A phase correlation method 
implemented in the MATLAB Image Processing Toolbox was chosen for the registration. 

In clinics, 3$\times$3- and \six\ are two most widely used scanning areas despite that \sixas\ usually presents inadequate image resolution. In order to mimic this real-life scenario, we trained all models only on the aforementioned LR-HR pairs of the overlapping foveal-central 3$\times$3-mm area. In total, 233 image pairs were used for training and 56 were reserved for validation. 
Notably, another 36 image groups, each of which consists of one foveal-centered plus four parafoveal-centered \threeas\ paired with their corresponding LR version from the \sixa, were used as the external test set, i.e., 180 image pairs in total.
Since ground truth images for the parafoveal regions are not routinely acquired in clinics, this dataset served as the test set only. 

\begin{table*}[!b]
\centering
\begin{tabular}{ccc}
\Xhline{1.5pt}\rule{0pt}{1.2em}    
                      & Foveal-central \three\ area                           & Whole \six\ area                                   \\ \cline{2-3} \rule{0pt}{1.2em}
                      & PSNR\up\ / SSIM\up\ / LPIPS\down                            & PSNR\up\ / SSIM\up\ / LPIPS\down           \\\hline \rule{0pt}{1.2em}
Frangi \cite{frangi1998}               & 15.191±0.783 / 0.489±0.065 / 0.230±0.024          & 15.209±0.592 / 0.514±0.049 / 0.227±0.040          \\
Gabor \cite{gabor2013}                 & 17.411±0.815 / 0.456±0.053 / 0.381±0.024          & 17.786±0.886 / 0.498±0.047 / 0.356±0.042          \\
SRCNN \cite{srcnn2014}                 & 17.074±0.715 / 0.402±0.047 / 0.492±0.023          & 17.630±0.897 / 0.514±0.048 / 0.462±0.050          \\
VDSR \cite{vdsr2016}                  & 18.485±0.927 / 0.560±0.056 / 0.267±0.025          & 18.228±0.962 / 0.573±0.050 / 0.252±0.041          \\
DRCN \cite{drcn}                  & 17.778±0.765 / 0.504±0.056 / 0.302±0.028          & 17.587±0.949 / 0.537±0.049 / 0.276±0.042          \\
MemNet \cite{memnet2017}                & 18.399±0.897 / 0.548±0.056 / 0.259±0.023          & 18.161±0.993 / 0.560±0.046 / 0.250±0.039          \\
EnhanceNet \cite{enhancenet2017}            & 17.663±0.962 / 0.505±0.061 / 0.305±0.028          & 17.996±0.951 / 0.544±0.050 / 0.277±0.043          \\
HARNet \cite{harnet2020reconstruction}                & 18.660±0.839 / 0.569±0.049 / 0.266±0.023          & 18.480±0.935 / 0.579±0.051 / 0.254±0.041          \\
TTSR \cite{ttsr}                  & 19.584±0.651 / 0.676±0.029 / 0.130±0.018          & 19.325±0.873 / 0.668±0.041 / 0.124±0.028          \\
\textbf{LTGNet(Ours)} & \textbf{19.815±0.622 / 0.691±0.027 / 0.122±0.013} & \textbf{19.538±0.884 / 0.682±0.040 / 0.117±0.027} \\ \Xhline{1.5pt}

\end{tabular}
\caption{PSNR/SSIM/LPIPS and the standard deviation results of different SR methods for the foveal-central \three\ area and the whole \six\ area. }
\label{table:stats}
\end{table*}
\subsection{Implementation Details}

The encoder included three blocks to extract features at different scales, which were composed of 1-2, 3-7, 8-12 layers of the Vgg19 \cite{vgg19simonyan2014very} network, respectively. It was initialized to the pre-trained weights on ImageNet \cite{imgnet} and fine-tuned during the training stage. We adopted the texture searching procedure described in TTSR \cite{ttsr} in the RefSR framework. Considering computational cost and the network capability, three MSFP blocks were stacked in the LTG. The Ref image was another randomly picked \threea, and Ref$\downarrow$ was the corresponding \sixa. Based on the validation experiments, $\lambda_{tg}$ was selected to be 0.3. $\lambda_{per}$ and $\lambda_{adv}$ were set to $1e-2$, $1e-3$, respectively. The learning rate for LTG, encoder, and the remaining components were set to $1e-4,\ 1e-6$ and $5e-5$. The learning rate was decayed by a factor of 0.7 for every 100 epochs, and the model was trained for 200 epochs in total. 

\subsection{Quantitative Results and Comparison}

We used peak-signal-to-noise ratio (PSNR), structural similarity (SSIM) and learned perceptual image patch similarity (LPIPS) to evaluate the model performance \cite{psnrssim,lpips}. Higher PSNR and SSIM and lower LPIPS values indicate better reconstruction outcomes. PSNR and SSIM are widely used in evaluation of image SR task and mainly focus on measuring the fidelity of the reconstructed images. On the other hand, LPIPS, which uses a pre-trained AlexNet \cite{alexnet2017} to extract features of the two images and calculates the difference between them, reflects visual quality that is closer to human perception. 

\begin{figure*}[!b]
    \centering
    \includegraphics[width=1\textwidth]{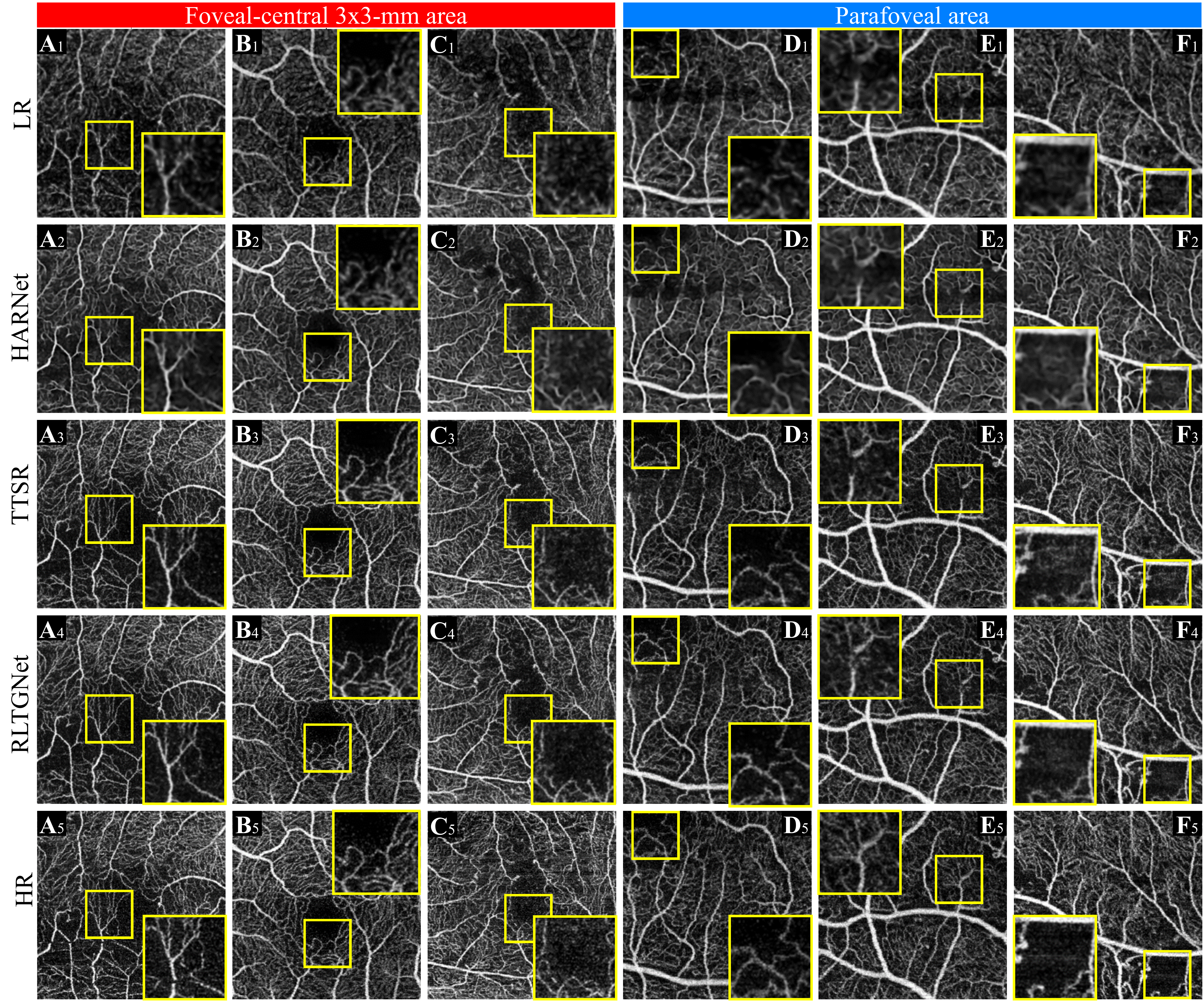}
	\caption{Qualitative results for foveal-central \three\ area and parafoveal area. (Zoomed in for clearer comparison.)}
    \label{fig:example}
\end{figure*}
Quantitative results of different methods are summarized in Table. \ref{table:stats}. All models were trained only with the overlapped foveal-central \three\ area to fit real-life constraints. The held-out validation set from the same region and the specially collected test set for the whole \six\ area were used to quantify the performance.  We can see that LTGNet has significantly better performance than state-of-the-art methods. Even without the Ref image at the testing stage, the LTG can generate useful textures automatically and outperforms TTSR \cite{ttsr}, a state-of-the-art RefSR model, by a considerable margin. In addition, our model achieves not only the best score for the foveal-central \three\ area but also for the whole \six\ area, demonstrating outstanding clinical reliability. Furthermore, the variance of our model performance is relatively small when compared with the other methods, showing great robustness. 

\subsection{Qualitative Results}
Exemplar images are shown in Fig. \ref{fig:example} for qualitative comparison. The regions of interest (ROIs) are denoted by yellow boxes, and the corresponding zoomed-in image patches are placed at the corner. We can see that the proposed LTGNet produced visually appealing images, which are almost indistinguishable by the human eye. It can also be observed that LTGNet outperformed other methods in terms of details reconstruction, noise suppression and correctness. 
As shown in Fig. \ref{fig:example} (A), (B), (D) and (E), LTGNet was more capable of handling subtle vessels reconstruction and preserving vessel connectivity. In addition, noise in the FAZ area was significantly reduced, as shown in Fig. \ref{fig:example} (C). As for the parafoveal area, even when the noise in the LR image was misleading, our method could identify the true vessels and recover the details in an accurate manner (Fig. \ref{fig:example} F), presenting fewer false signals when compared with the other methods. 
\begin{figure*}[t]
    \centering
    \includegraphics[width=0.9\textwidth]{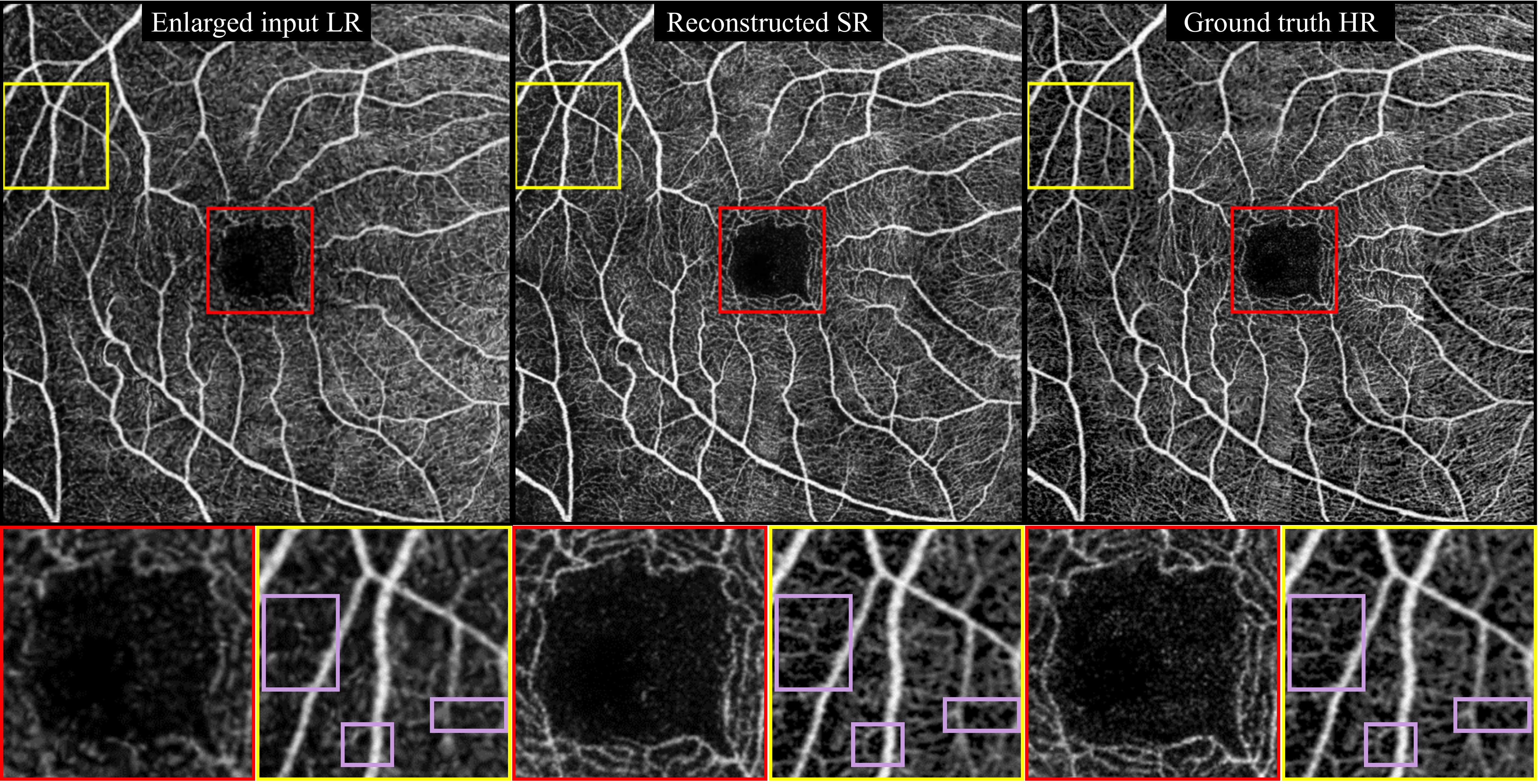}
	\caption{Reconstructed results on the whole \six\ area.}
    \label{whole}
\end{figure*}
Fig. \ref{whole} (B) shows a reconstructed whole \six\ OCT angiogram. The enlarged LR input shown in Fig. \ref{whole} (A) was obtained by upsampling the raw \sixa\ by a factor of 2 via bicubic interpolation so that every pixel in it had the same physical scale as the high-resolution \threea. The ROIs are labeled by colored boxes, and the zoomed-in patches are displayed in the next row for better visual comparison. It is noticeable that the microvasculature (denoted by purple boxes) was reconstructed quite precisely. Moreover, the noises present in the FAZ area was significantly less than in the directly acquired \six\ HR image, while the FAZ was still clearly delineated.


\subsection{Sensitivity Analysis of Hyperparameters}


\begin{figure*}[!b]
    \centering
     \subfloat[Foveal-central \three\ area]{
     \includegraphics[width=0.48\textwidth]{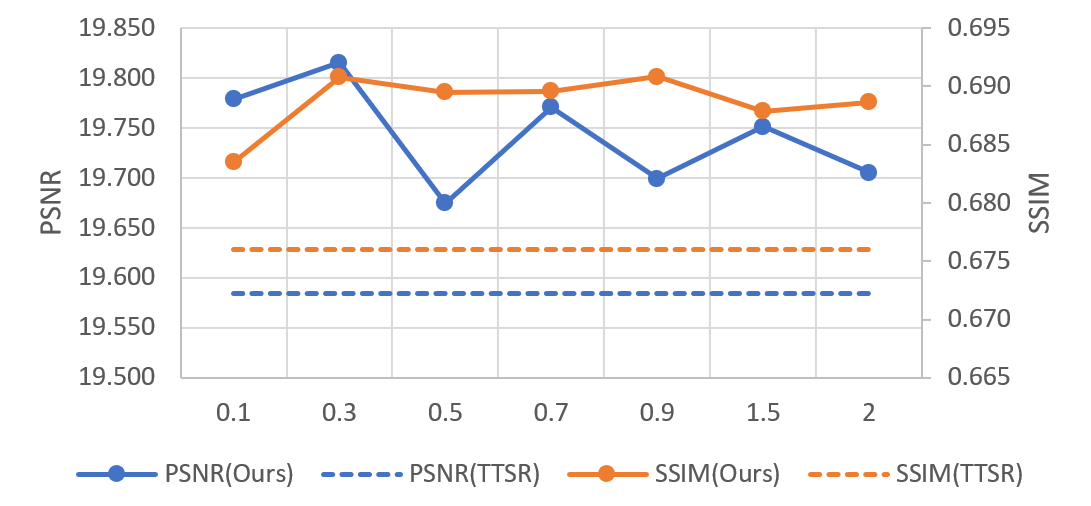}
     }
     \hfill
     \subfloat[Whole \six\ area]{
     \includegraphics[width=0.48\textwidth]{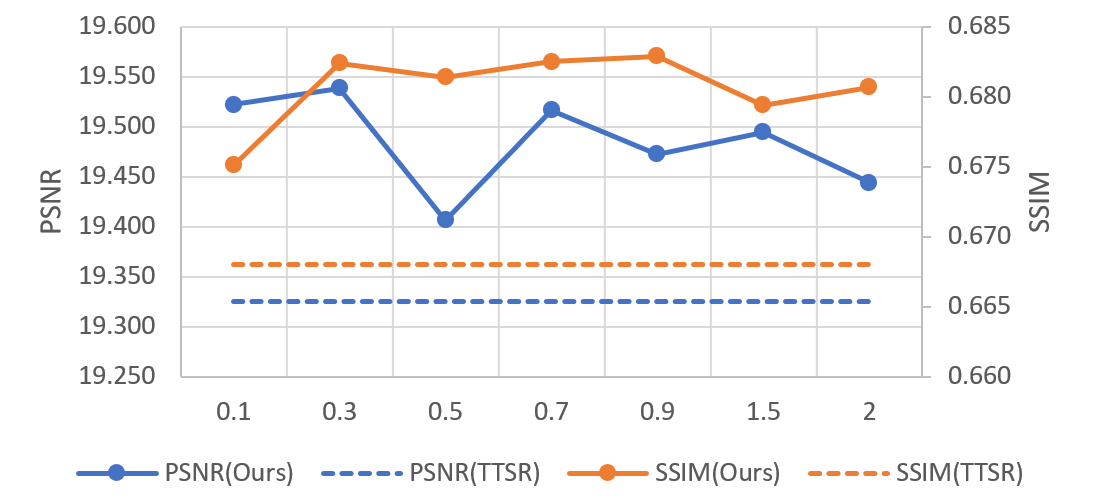}
     }
     \hfill
     \label{fig:abl}
     \caption{Model's performance change with $\lambda_{tg}$. The dotted lines are the results of TTSR \cite{ttsr}. }
     \label{fig:abl}
\end{figure*}
The performance change of the model when varying the hyper-parameter $\lambda_{tg}$ is shown in Fig. \ref{fig:abl}. 
It can be observed that our model robustly outperformed the TTSR when $\lambda_{tg}$ is between 0.1 and 2, demonstrating the innate superiority of our model. 
We also conducted extra experiments to see the model's performance change when varying the number of MSFP blocks ($m$) stacked in the LTG, with $\lambda_{tg}$ fixed to 0.3. The results and the model complexity are shown in Table. \ref{table:msfp}. Surprisingly, we found that even with only \emph{one} MSFP block, our model already surpassed TTSR. According to the performance on the validation set, the model was finalized to include three MSPF blocks in the LTG with the considerations of model capacity and computational costs. 
\begin{table}[htp]
\centering
\begin{tabular}{cccccc}
\Xhline{1.5pt}\rule{0pt}{1.2em}
              & \multicolumn{2}{c}{Foveal-cener 3x3 area} & \multicolumn{2}{c}{Whole 6x6 area} &                  \\ \cline{2-6} \rule{0pt}{1.2em}
\textit{m}    & PSNR                 & SSIM               & PSNR             & SSIM            & \#params        \\ \rule{0pt}{1.2em}
1 & 19.806 & 0.691 & 19.402 & 0.676 & 4196673 \\
2 & 19.792 & 0.691 & 19.529 & 0.683 & 4680769 \\
3 & 19.815 & 0.691 & 19.538 & 0.682 & 5164865 \\
4 & 19.753 & 0.690 & 19.457 & 0.682 & 5648961 \\
5 & 19.769 & 0.691 & 19.532 & 0.683 & 6133057 \\
\textit{TTSR} & \textit{19.584}      & \textit{0.676}     & \textit{19.325}  & \textit{0.668}  & \textit{3589313} \\ \Xhline{1.5pt}
\end{tabular}
\caption{Model performance and number of parameters when varying the number of MSFP blocks.}
\label{table:msfp}
\end{table}

\subsection{Ablation Study}
\begin{figure}[!b]
    \centering
    \includegraphics[width=0.48\textwidth]{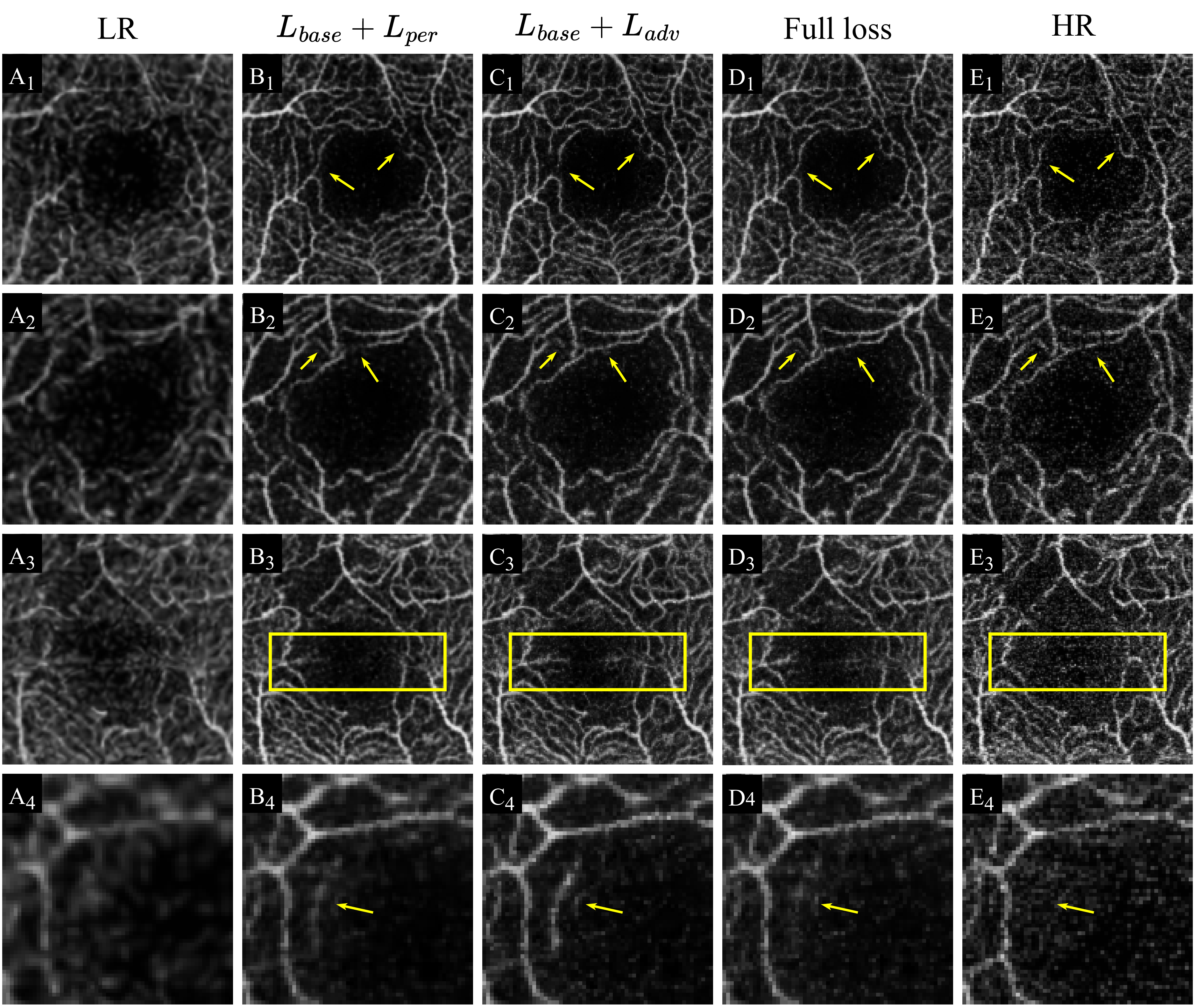}
	\caption{Selected examples to demonstrate the effectiveness of $\loss_{per}$ and $\loss_{adv}$. }
    \label{img:loss}
\end{figure}

We investigated the influence of different components in the loss function, and the corresponding results are shown in Table. \ref{table:loss}. We can see that removing $\loss_{ref}$ brings catastrophic degradation, verifying the necessity of guiding the texture generation module during the training process. 
\begin{table}[htpb]
\centering
\begin{tabular}{ccccc}
\Xhline{1.5pt}\rule{0pt}{1.2em}
$\loss_{ref}$  & $\loss_{per}$ & $\loss_{adv}$ &  PSNR & SSIM\\\hline \rule{0pt}{1.2em}
\tikzxmark &   &   & 19.176 & 0.624 \\
  & \tikzxmark &   & 19.720 & 0.692 \\
  &   & \tikzxmark & 20.013 & 0.692 \\
 \rowcolor[gray]{0.8} & & & 19.815 & 0.691\\ \Xhline{1.5pt}
\end{tabular}
\caption{Change of model performance when excluding different parts of the loss function on the validation set. The model trained with full loss is highlighted in gray. } 
\label{table:loss}
\end{table}
Introducing $\loss_{per}$ and $\loss_{adv}$ brings some degradation in quantitative results but can benefit visual quality, which is well recognized by current literature \cite{perceptual2017enhancenet,perceptual2019textimg,perceptuallucas2019,ttsr,ganloss2022review}.
Fig. \ref{img:loss} helps understand the visual improvements brought by $\loss_{per}$ and $\loss_{adv}$. 
We packed $\loss_{rec}$ and $\loss_{ref}$ as $\loss_{base}$ and trained the model with $\loss_{base}+\loss_{per}$,  $\loss_{base}+\loss_{adv}$, $\loss_{base}+\loss_{per}+\loss_{adv}$ (i.e., full loss), respectively, and investigated the visual quality of these results. Some selected examples are displayed in Fig. \ref{img:loss}. For the first row, we can see that vessel signals in the images generated by the model trained with $\loss_{adv}$ (C$_1$) are stronger than the one without it (B$_1$). However, such an "aggressive" reconstruction manner could lead to undesirable results. 

For example, some weak vessel signals could be mistakenly removed (pointed by the arrows in C$_1$). Contrastingly, these weak signals are better preserved in B$_1$ despite the relatively low strength. By combining $\loss_{per}$ and $\loss_{adv}$, the weak vessel signals are well preserved and presented clearly (D$_1$). Similar results can be observed in the second row, where D$_2$ presents both good sides of $B_2$ and $C_2$. As for the model trained with $\loss_{base}+\loss_{per}$, it tends to do the reconstruction in a more conservative manner by only enhancing the signals of strong confidence. This could lead to a relatively low signal strength but can help with the correction of falsely constructed vessels. Comparing B$_3$ and C$_3$, we can see that the false signals are much weaker in B$_3$. As for D$_3$, the false signals are weakened greatly, showing that including $\loss_{per}$ is also beneficial to reduce the side effects brought by $\loss_{adv}$. Moreover, we observed some surprising benefits when combining these two components as well. As shown in the fourth row, the false signal in D$_4$ is the weakest one among B$_4$, C$_4$ and D$_4$. Therefore, it can be observed that each component in the loss function plays an indispensable role. 
\begin{figure*}[t]
    \centering
    \includegraphics[width=0.9\textwidth]{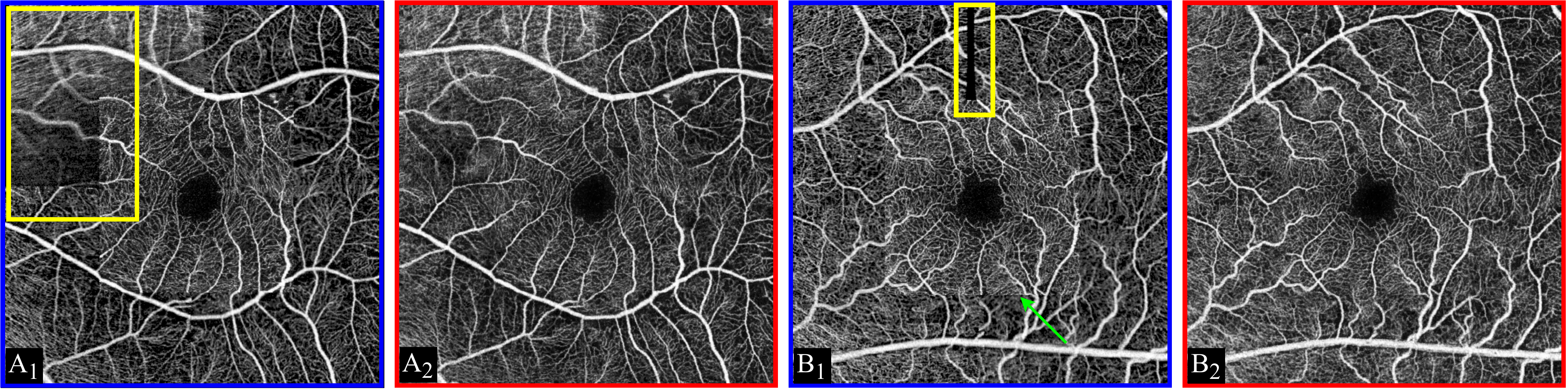}
	\caption{Comparison between the HR \six\ montage and the reconstructed results (denoted by blue and red boxes, respectively). The ROIs are highlighted by yellow boxes.}
    \label{low_quali}
\end{figure*}
\section{Discussion}

OCT angiograms with large FOVs are helpful in retinal disease assessment since they can provide comprehensive information about retinal conditions. For instance, important features like non-perfusion may present in the parafoveal areas, which can not be detected by only looking into the standard \threeas. Therefore, efforts have been put into obtaining HR OCT angiograms with large FOVs via upgrading hardware or applying specific acquisition practices.
Improving A-line rate is one representative approach via hardware upgrading, but the equipment is expensive and may harm the vascular sensitivity \cite{octasargan2022}. It has also been proposed to solve the undersampling problem by extending the acquisition time so that the resultant angiogams have the same scanning density as those with small FOVs. This approach places a larger burden on patients' tolerance and is even more challenging for the elderly and children, making it clinically impractical. \cite{octa3} proposed to break down the whole task into five subtasks, namely, obtaining five standard \threeas\ for the foveal-central \three\ and parafoveal areas. After that, the five \threeas\ are stitched together to form an HR \sixa. We adopted this approach to collect our independent test set. However, we encountered several problems during data collection and preprocessing. Firstly, the patients were required to look in a specific direction during the acquisition of parafoveal-centered angiograms. This turned out to be too demanding for the patients, which introduced extra artifacts. One typical example is shown in Fig. \ref{low_quali} A$_1$. In addition, it was also difficult to stitch the five angiograms to form a perfect montage. Common artifacts included the gap between the images (denoted by the yellow box in Fig. \ref{low_quali} B$_1$), abrupt changes in brightness between the foveal and parafoveal areas and imperfect stitching edges (pointed to by the green arrow in Fig. \ref{low_quali} B$_1$). In contrast, the reconstructed images present smoother brightness distribution and fewer destructive artifacts. Therefore, image SR techniques are a more convenient and effective way to obtain OCT angiograms with large FOVs as well as high resolution. 

Among the SR methods, filter-based approaches cannot accurately handle the complex vasculature and the great details presented in the OCT angiograms, which often results in over-smoothing and loss of capillaries \cite{harnet2020reconstruction, octasargan2022}. By contrast, learning-based approaches have shown extraordinary power in handling complicated computer vision tasks, including segmentation, denoising, and super-resolution, to name a few \cite{dlseg2020, dldenoise2022, dlsr2021}. RefSR is a novel class of learning-based SR methods, which utilizes information from an extra high-resolution image (Ref image) to assist the SR process. With such information from the target domain, the model performance is further boosted. 
However, the available information is limited to the input Ref image, which is still suboptimal. In our method, we built a LTG to learn textures contained in the training samples and let it generate useful textures during inference according to the input. In this way, the available texture space can be enlarged, and experimental results verify the effectiveness of our design.  
Another drawback of RefSR is its sensitivity to the selection of Ref images. To be concrete, good Ref images containing similar contents might help improve the reconstruction result to a great extent, but misleading Ref images might harm the performance.
Thus, we aimed to build a subnet to generate textures automatically according to the input. This avoids the tricky selection process, and more robust performance can be achieved. Extensive experimental results described in the quantitative and qualitative results prove that this design is more powerful than randomly choosing Ref images. To verify the superiority of our method, we performed extra experiments in which we shuffled the Ref images in the test set with different random seeds to alleviate the side-effect of randomness. It was found that all combinations show worse performance than our proposed method (Table. \ref{shuffle}), verifying the intrinsic advantage of our design.
\begin{table}[htpb]
\centering
\includegraphics[width=0.48\textwidth]{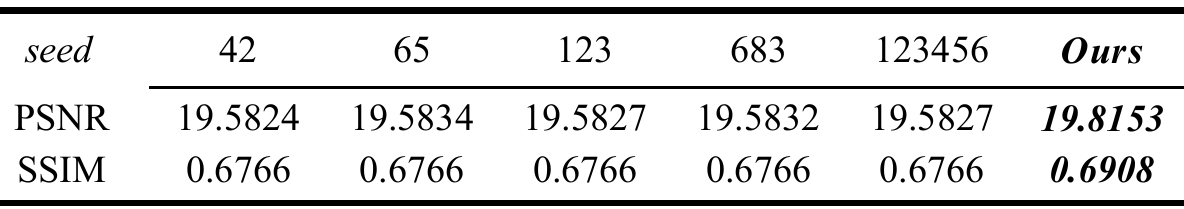}
\caption{Performance of TTSR when testing with different random seeds.}
\label{shuffle}
\end{table}

\begin{figure}[htpb]
    \centering
    \includegraphics[width=0.45\textwidth]{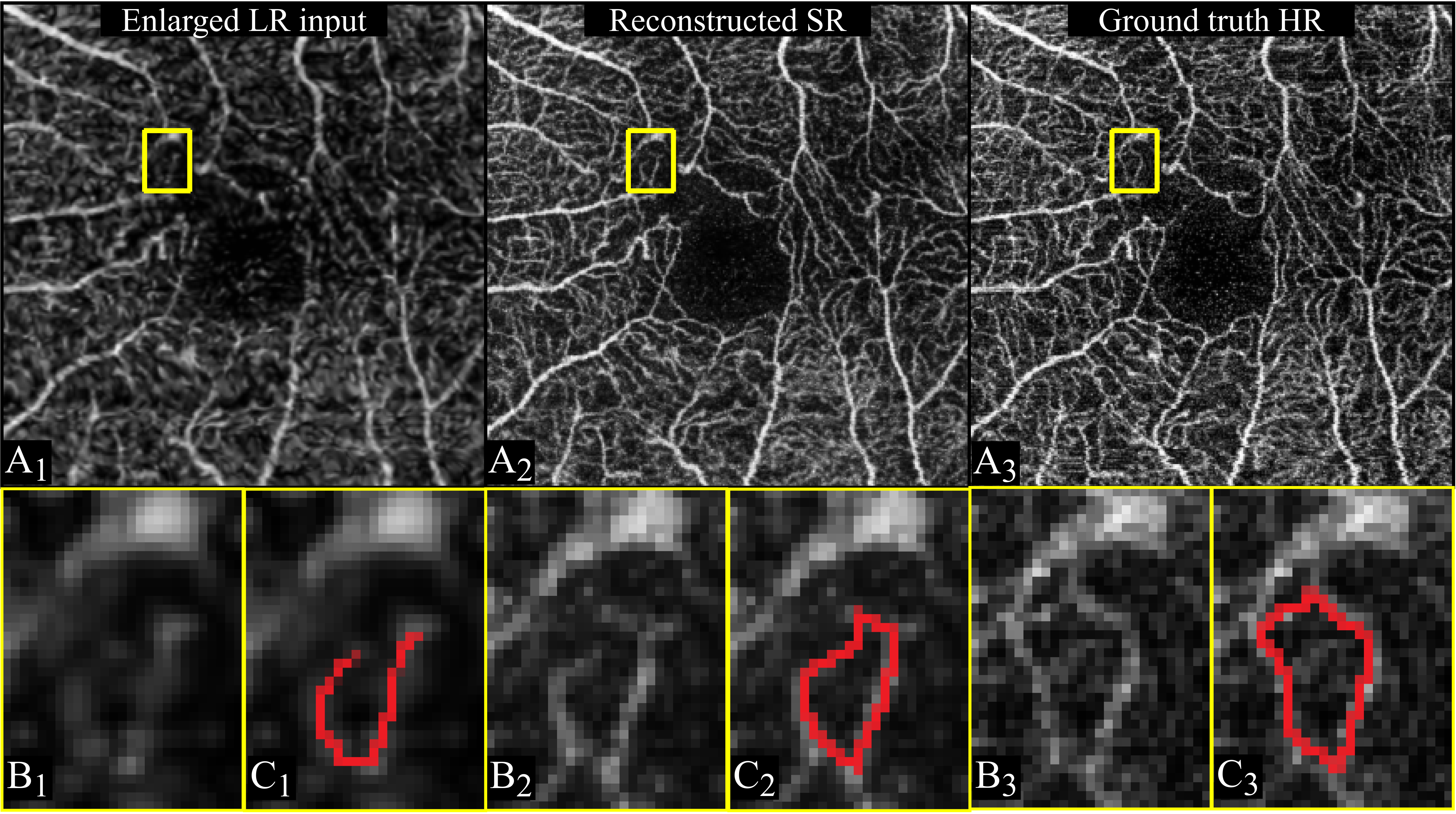}
	\caption{Sample discrepancy in LR and HR images.}
    \label{discrepency}
\end{figure}

During this study, we observed some interesting phenomena. When looking into the details of the HR ground truth and the SR results, some structural differences were recognized. 
In Fig. \ref{discrepency}, it is noticeable that the shape of the small vessel ring in the yellow box is quite different between LR and HR. The zoomed-in patches are shown in B$_1$ to B$_3$, and shape of the vessel ring is annotated accordingly in C$_1$ to C$_3$. It can be seen that the reconstructed result is very close to the indication of the signals in the LR image, while it is quite dissimilar to the HR one. This indicates that the LR and HR imags are not perfectly aligned. This discrepancy may hinder the model from further improving the quantitative results. This also gives a reason as to why the PSNR and SSIM results of the reconstructed images (about 20 and 0.7) are drastically lower than those from normal natural image SR tasks, where state-of-the-art methods usually achieve over 30 in PSNR and 0.9 in SSIM when dealing with $\times 2$ upscaling. We infer the reason to be that the structure of the retina is slightly different during the acquisition of the 3$\times$3- and \sixas\ because of the time difference between taking the two images. This is reasonable since organ changes happen frequently in living organisms. In the future, we may come up with an approach that could explicitly reflect clinical significance to evaluate the quality of the reconstructed results. For example, we could identify the capillary structural difference and more heavily penalize on significant changes that may be misleading for diagnosis, like mistakenly broken vessels or falsely constructed microaneurysms, while placing a less penalty on inconsiderable ones. 

\section{Conclusion}
OCT angiograms with large FOVs are useful for retinal disease assessment but are usually undersampled in current commercial OCTA systems. Despite recent progress, obtaining large-FOV OCT angiograms with sufficient resolution is still challenging, and using image SR techniques remains the optimal solution to generate HR large-FOV OCT angiograms under common clinical equipment conditions and constraints like patients' tolerance. Therefore, we present a novel RefSR model called LTGNet for OCT angiogram SR task in this paper. It consists of a conventional RefSR model along with a learnable texture generator (LTG). With the LTG, the model can generate textures according to the LR input, and it does not require any reference images at inference time, making it invulnerable to the selection of Ref images. Furthermore, experimental results show that although LTGNet is only trained on the foveal-central \three\ area to mimic real-life constraints, its performance on the whole \six\ area is still robust and satisfactory. More importantly, our model demonstrates good robustness and capability when varying the hyperparameters in the settings, showing the innate superiority of the design. 
Conclusively, our work helps mitigate the quality degradation problem when increasing the scanning area in OCTA imaging to a great extent.
In the future, we will focus on quantifying and improving the model performance in terms of diagnostic significance.


\bibliographystyle{ieeetr}
\bibliography{refs.bib}



 





\end{document}